\documentstyle[epsf]{mn}
\newcounter{saveeqna}

\newcounter{saveeqnb}
\newcommand{\alpheqntext}
{\setcounter{saveeqnb}{\value{equation}}%
\stepcounter{saveeqnb}%
\setcounter{equation}{0}%
\renewcommand{\theequation}
{\mbox{\arabic{saveeqnb}.\alph{equation}}}}
\newcommand{\reseteqntext}
{\setcounter{equation}{\value{saveeqnb}}%
\renewcommand{\theequation}{\arabic{equation}}}
\newcommand{\msun}{\mbox{$\rm {M_{\odot}}$}}

\newcommand{\mbh}{\mbox{$M_{\rm{BH}}$}}
\newcommand{\kms}{\mbox{$\rm{kms^{-1}}$}}
\newcommand{\kpc}{\mbox{$\rm{kpc}$}}
\newcommand{\beq}{\begin{equation}}
\newcommand{\eeq}{\end{equation}}

\newcommand{\vv}{\mbox{$\bf v$}}

\newcommand{\sig}{\:\lower0.6ex\hbox{$\stackrel{\textstyle >}{\sim}$}\:}
\newcommand{\sil}{\:\lower0.6ex\hbox{$\stackrel{\textstyle <}{\sim}$}\:}
\newcommand{\sigs}{\:\lower0.4ex\hbox{$\stackrel{\scriptstyle
      >}{\scriptstyle \sim}$}\,}
\newcommand{\sils}{\:\lower0.4ex\hbox{$\stackrel{\scriptstyle
      <}{\scriptstyle \sim}$}\,}
\begin{document}
%
\title[Constraints on massive black holes as dark matter candidates]
{Constraints on massive black
holes as dark matter candidates using Galactic globular clusters }

\author[R. Klessen and A. Burkert]
{Ralf Klessen$^1$ and Andreas Burkert$^2$\\
Max-Planck-Institut f\"ur Astrophysik, K\"onigstuhl 17, 69117
Heidelberg, Germany\\
$^1$klessen@mpia-hd.mpg.de\\
$^2$burkert@mpia-hd.mpg.de}

\maketitle
\begin{abstract}
This work considers the idea of
massive black holes being the constituents of the Galactic dark matter halo.
It constrains the maximum black hole mass to $\mbh \sil 5 \times 10^4
\msun$ by examining their
influence on the population of globular clusters in our Milky Way. In
the adopted halo model, globular clusters are  exposed to constant
bombardment of halo objects on their
orbits through the Galaxy and thus  will steadily gain internal
energy.
Depending on the mass of these halo objects and the structural
parameters of the globular clusters, they
 can be disrupted on time scales of a few
billion years and below.
These disruption time scales are calculated using a modification of the
well known (classical) impulsive approximation and compared with direct
N-body simulations of such encounter events to ensure the method works
correctly.
For a set of ten prototypical globular cluster models and black hole
masses ranging from $10^3$ to $10^7\:\msun$,
Monte-Carlo-simulations of $10\,000$
encounter histories over the period of 10 billion years were
calculated each, at three different galactocentric distances \mbox{$R =
5$}, $10$ and $15\, \kpc$. These
data were compared with the real globular cluster population in our
Galaxy and used to obtain the above constraint of $\mbh \sil 5 \times
10^4 \msun$.
\end{abstract}


\section{Introduction}
Comparison of dynamical and photometrical mass determinations in
galaxies like our own
 strongly indicates the existence of a non-luminous dark matter
 component, that accounts for most of the mass in these galaxies. The
 constancy of $\rm{HI}$-rotation curves even far beyond the
 optical edge of galactic disks
furthermore indicates a $1/r^2$-density distribution of this dark
component (Rubin~et.~al.~1982, 1985). Thus one  question naturally arises:
What is the nature of this dark component?
A whole variety of dark matter candidates have been proposed in the
literature, ranging from light leptonic particles to  baryonic gaseous matter
(possibly in the  form of cold molecular clouds with fractal structure;
Pfenninger~et.~al.~1994a,b) to low mass stars such as white dwarfs
or ``Jupiters''   and finally black holes with masses $\mbh
\simeq 10^6 \msun$ as remnants of a
	population of very massive primordial stars
(for a brief overview see e.g. Chap.~18 in Peebles~1993
and references therein).

\begin{figure*}
\begin{minipage}[t]{16cm}
\begin{minipage}[t]{7.5cm}
\epsfxsize=7.5cm \epsfbox{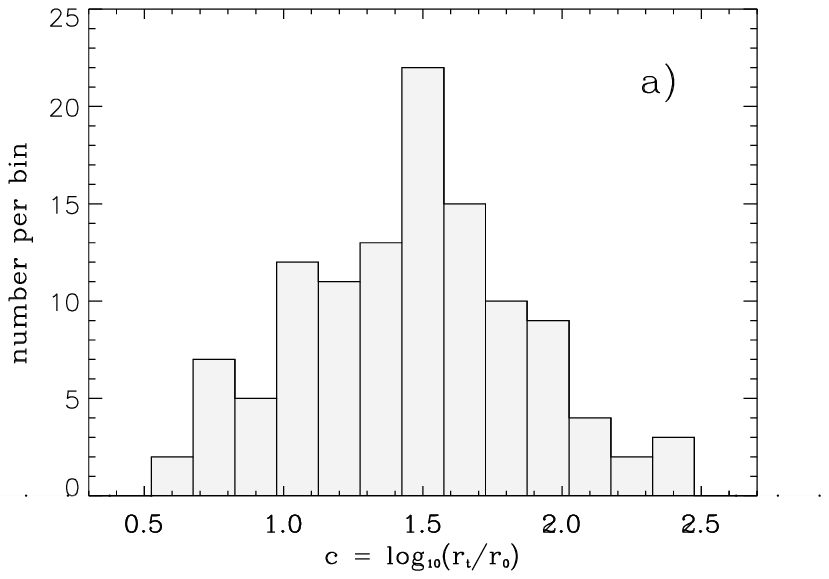}
\end{minipage} \hspace{0.2cm}
\begin{minipage}[t]{7.5cm}
\epsfxsize=7.5cm \epsfbox{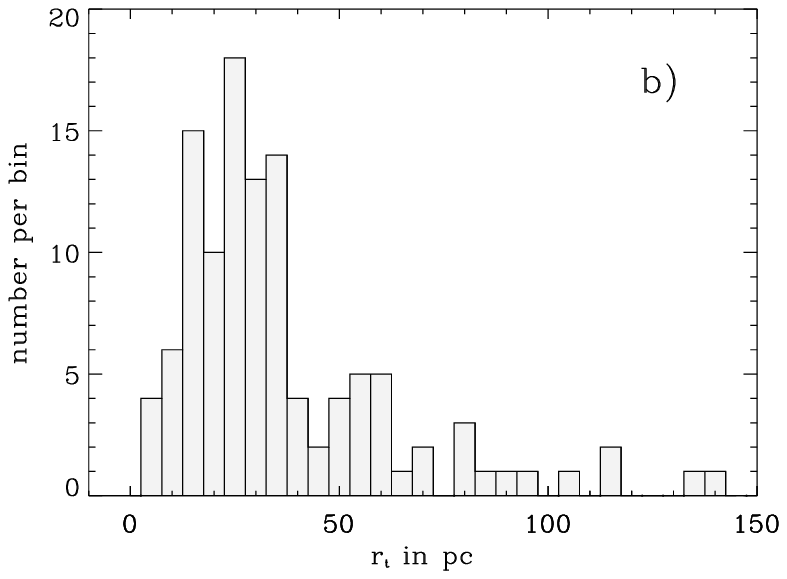}
\end{minipage} \hfill
\vspace{0.2cm}
\begin{minipage}[t]{7.5cm}
\epsfxsize=7.5cm \epsfbox{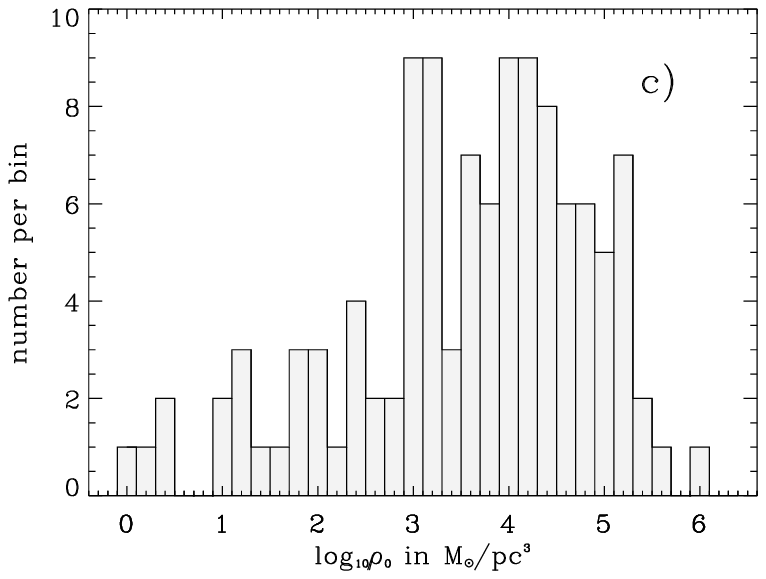}
\end{minipage} \hspace{0.2cm}
\begin{minipage}[t]{7.5cm}
\epsfxsize=7.5cm \epsfbox{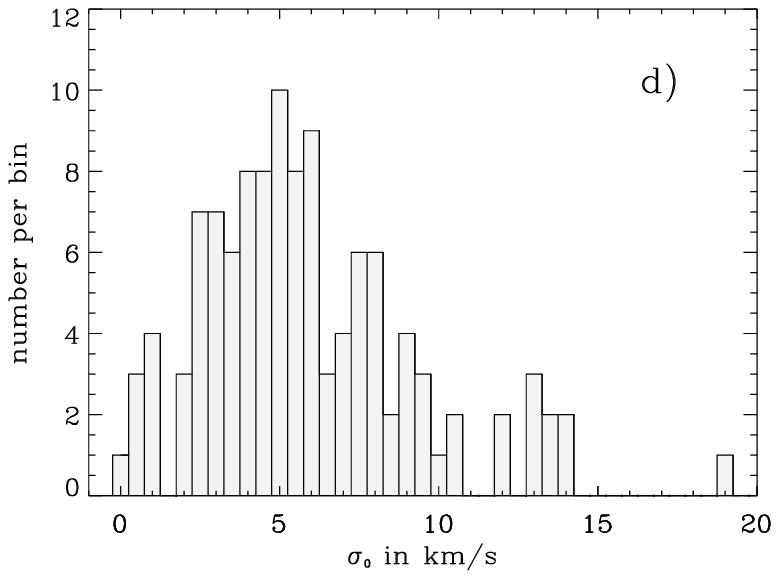}
\end{minipage} \hfill
\end{minipage}
\caption{
\label{king-parameter-pic}
Distribution of a) concentration index $c$, b) tidal radius $r_{\rm t}$, c)
logarithm of the central star density $\log_{10} \rho_0$ and d) central
velocity dispersion $\sigma_0$ for 134 globular clusters in the Milky
Way
(data from Webbink~1985).}
\end{figure*}

\begin{figure*}
\begin{minipage}[t]{16cm}
\epsfxsize=16cm \epsfbox{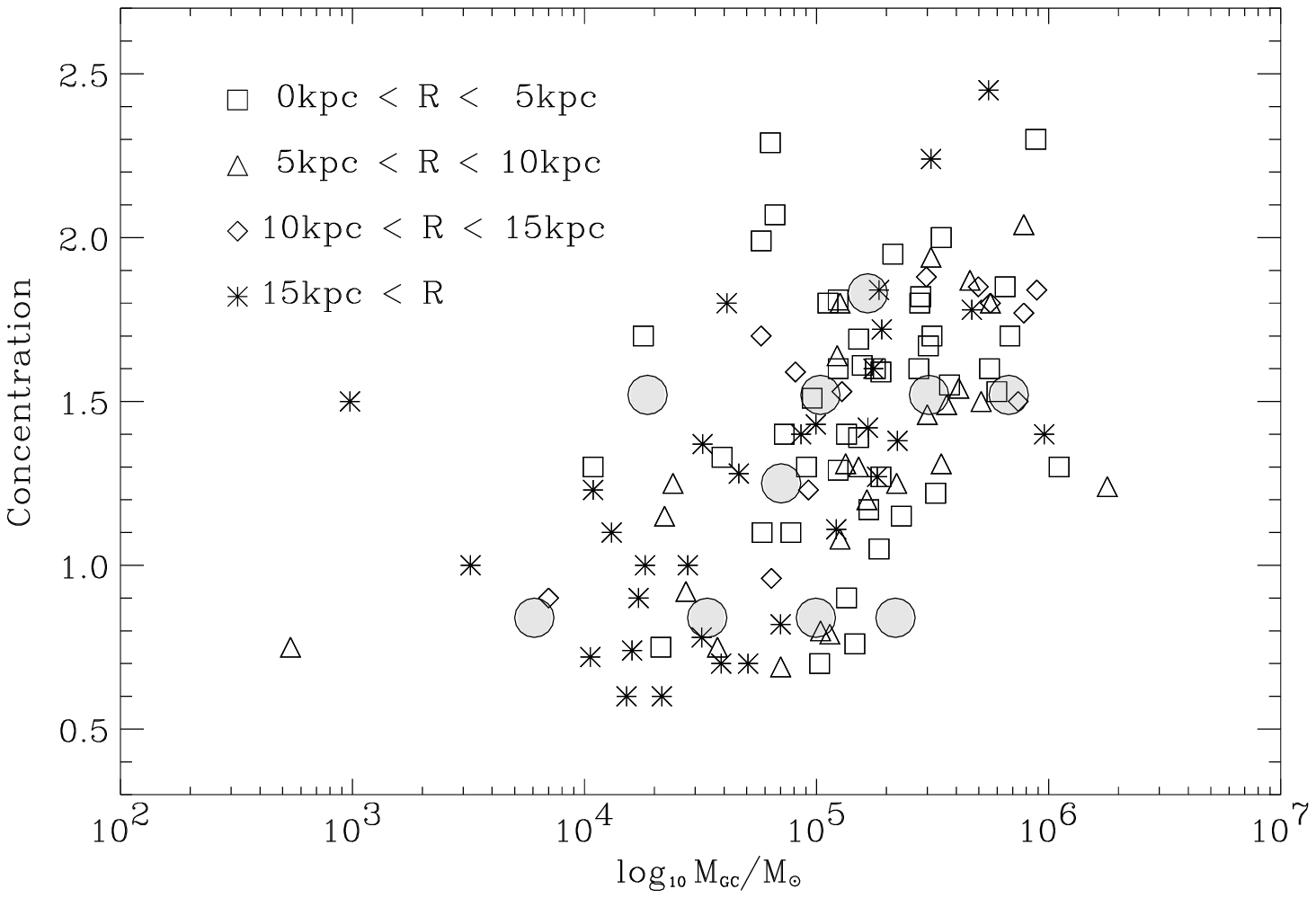}
\end{minipage}
\caption[]{
\label{fig-conc-mass}
Distribution of total mass $M_{\rm GC}$ and concentration index $c$
for 113 Galactic globular clusters of the Webbink~(1985). The
shaded circles are the $M_{\rm GC}$ and $c$ values of the ten model
clusters in our simulations.
}
\end{figure*}

The latter are of theoretical interest, since the Jeans mass at high
red shift (just after the epoch of recombination) was of the order of
$10^6\,\msun$ (Dicke~\&~Peebles~1968) and black holes with this
mass could very effectively heat an initially cold disk in our Galaxy
explaining the observed properties (Lacey~\&~Ostriker~1985). But there
also are many arguments against massive black holes as main constituents of
the Galactic dark halo. Hut~\&~Rees (1992) argued that black
holes with $~10^6\,\msun$ in the inner few kiloparsecs would sink into
the Galactic center within a fraction of the Hubble time due to
dynamical friction. This process would lead to  consequences that strongly
disagree with observations. Furthermore massive black holes would on
passages through the Galactic disk
accrete interstellar matter and thus violate background light
constraints (Carr~1979). And massive black holes would finally cause
gravitational lensing of cosmologically distant sources.
Garrett~et~al.~(1994) studied this
effect on $\lambda\,18\:\rm{cm}$ VLBI maps of the gravitational
lens system $0957+561\,\rm{A,B}$ and conclude that black holes
with $\mbh > 3 \times 10^6\,h^{-1}\,\msun$ cannot account for the dark
matter in the halo of the main lensing galaxy.
Rix~\&~Lake (1993)  again applied
the disk heating scenario to dwarf galaxies of the Local Group and
derived an upper limit of $\mbh \sil 10^4 \,\msun$.
On the other hand, Carr, Bond \& Arnett (1984) derived a strong
lower limit of $\mbh > 10^2\,\msun$  by examining the possible cosmological
consequences of population~III~stars. Their black hole remnants must
exceed this limit, otherwise the enrichment of the interstellar medium
would be too pronounced.

This paper facilitates a different approach: it uses the fragility of
globular clusters in our Galaxy to constrain \mbh.
Adopting the hypothesis that the dark halo consists entirely of massive black
holes, globular clusters are
subject to constant bombardment of these
objects on their orbits through the Milky Way.
 The observed properties of globular clusters thus
set strong limits on the mass distribution of these black holes.  To
calculate the response of globular clusters to encounters with massive
black holes we use a modification of the impulsive approximation
  (Spitzer~1958) and test this analytical approach by direct
N-body simulations of such an event. The excellent agreement of both
methods shows the validity of the assumptions of the analytical
approach and allows us to study a large region in parameter space since
this method consumes much less  computer time. We
obtain analytical values for the average internal energy gain of globular
clusters due to the interaction with halo objects in the Milky Way.
Since this energy gain is mostly determined by only few central
encounters, the scatter around its mean value is very large.
To account for that scatter and to be able to make  statistically
sound predictions, we adopt ten model clusters and compute for every one of
them $10\,000$ encounter histories each at three different galactocentric
radii ($R=5$, $10$ and $15\:\kpc$)
and for a black hole mass range between $10^3$ and $10^7\:\msun$,
applying Monte Carlo methods.
If the increase of internal kinetic energy of the
system exceeds its binding energy, we assume it dissolves. Thus we
obtain  the fraction of globular clusters, which would
survive for $10^{10}$ years, depending
 on internal
parameters, such as concentration and total mass,
their galactocentric distance and the assumed
black hole mass. Comparing that data with observations we constrain
the maximum black hole mass to $M_{\rm BH} = 5 \times 10^4\:\msun$.

This value agrees with  the mass range obtained by other authors using
a variety of methods. It especially has to be compared with the
numbers calculated by Moore~(1993). He used similar arguments than
this paper applies, but did not take the huge spread in internal
energy gain for different encounter histories into account. He
calculated  for nine different globular
clusters in the outer Galactic halo one encounter history
each. Therefore his estimate is highly uncertain.
Taking into account different encounter histories of individual
globular clusters, the present work stands
statistically on more solid ground than previous work, leading to a
more precise estimate for the maximal allowed mass of compact dark halo
objects.

\section{Galactic Globular Clusters}
\label{globulars-properties}

Galactic globular clusters can very well be fitted by a King density
profile (King~1966, Illingworth~\&~Illingworth~1976).  King
models are lowered nonsingular isothermal spheres, which
are generated from the distribution function of
an  isothermal sphere by introducing an energy cut-off .
This results in a radial
cut-off at the  tidal radius $r_{\rm t}$.  King models form a
sequence that can be parametrized by the  concentration index $c \equiv
\log_{10}(r_{\rm t}/r_0)$.
 The core radius $r_0$ is determined
by the central density $\rho_0$ and the central velocity dispersion
$\sigma_0$:
\begin{equation}
r_0 \equiv \sqrt{ \frac{9 \sigma_0 ^2}{4 \pi G \rho_0}}
\label{umpf_umpf}
\end{equation}
A compilation of these parameters for 134 galactic globular clusters
from the Webbink~(1985) data set is shown in
Fig.~\ref{king-parameter-pic}.

According to this data we generated ten model clusters with different
parameter combinations (see Tab.~\ref{gc-models}) to simulate
the influence of massive compact halo objects on globular clusters of
different sizes and concentrations. Models \#0 to \#3 have average
concentration ($c = 1.52$) and vary in their central dispersion from
$\sigma_0 = 3.0 \kms$ to $9.0 \kms$. Models \#4 to \#7 are very little
concentrated ($c=0.84$) and have the same variation in $\sigma_0$ as above.
Due to their lower concentration their total mass $M$ is smaller by a
factor of roughly one third.
The sequence \#5, \#8 and \#9 ranges  from small to large concentrations
and has fixed $\sigma_0$.  Model \#1 represents the ``mean'' of
the Galactic globular cluster system and will be called the {\em
standard cluster}
throughout this paper. For the numerical calculations, each of these
clusters is represented by $25\,000$ test particles of equal mass.

Plotting concentration index $c$ against total mass $M$,
Fig.~\ref{fig-conc-mass} depicts that the Galactic population of
globular clusters is fairly well sampled by the adopted set of model
clusters (which are represented by the gray shaded circles).
However this is not true for the central density $\rho_0$, which
 in reality spans a range from $~10$ to $10^6\:\msun/\rm{pc}^3$
(see Fig.~\ref{king-parameter-pic}.c), whereas the model clusters have $\rho_0
\simeq 5 \times 10^3 \: \msun/\rm{pc}^3$. This has to be taken into account in
the later analysis,
when comparing the numerical results with observations.
 Figure~\ref{fig-conc-mass} also indicates an
interesting mass-distance relationship. Globular clusters at smaller
galactocentric radii tend to be more massive than more distant
ones. Due to equation~\ref{umpf_umpf} this holds also for $\rho_0$ and
$\sigma$: The inner globular clusters have on average higher central
densities and larger velocity dispersions. Closer to the Galactic
center, these globular clusters are more severely influenced by tidal
shocking during passage through the Galactic disk (and bulge in the
inner few kpc) and encounters with massive halo objects.
Therefore they  have to be more massive and sturdy to survive for a Hubble
time.

\begin{table}
\caption{
\label{gc-models}
Properties of the ten model clusters used in the simulations. The
parameters listed are central concentration index $c$, total mass
$M$, core radius $r_0$, tidal radius $r_t$, velocity dispersion
$\sigma_0$ and density $\rho_0$ in the central region of the clusters.
}
\begin{center}
\renewcommand{\arraystretch}{1.5}
\begin{tabular}[t] {|l|c|c|c|c|} \hline
model &  \#0 &  \#1 &   \#2 &   \#3 \\
\hline
$c=\log_{10}(r_{\rm t}/r_0)$ & 1.53 & 1.53 & 1.53 &  1.53 \\
$M$ in $10^5\,\msun$ & $0.19$ & $1.04$ & $3.05$ & $6.73$ \\
$r_0$ in $\rm pc$ & $0.5 $&$1.0$&$1.5$&$2.0$\\
$r_{\rm t}$ in $\rm pc$ & 16.8 & 33.7 & 50.6 & 67.4 \\
$\sigma_0$ in $\kms$ & 3.0 & 5.0 & 7.0 & 9.0 \\
$\rho_0$ in $10^3 \,\msun \rm{pc}^{-3} $&$ 6.0 $ &$ 4.1 $&$ 3.6 $&$ 3.3 $ \\
\hline
\hline
model &  \#4 &  \#5 &   \#6 &   \#7 \\
\hline
$c=\log_{10}(r_{\rm t}/r_0)$ & 0.84 & 0.84 &  0.84 &  0.84 \\
$M$ in $10^5\,\msun$ & $0.06$ & $0.34$ & $0.99$ & $2.19$ \\
$r_0$ in $\rm pc$ & $0.5 $&$1.0$&$1.5$&$2.0$\\
$r_{\rm t}$ in $\rm pc$ & 3.46 & 6.92 & 10.4 & 13.8 \\
$\sigma_0$ in $\kms$ & 3.0 & 5.0 & 7.0 & 9.0 \\
$\rho_0$ in $10^3\,\msun \rm{pc}^{-3} $&$ 6.0  $&$ 4.1 $&$ 3.6  $&$ 3.3 $ \\
\hline
\hline
model &  \#5 &  \#8 &   \#9 &  \\
\hline
$c=\log_{10}(r_{\rm t}/r_0)$ &0.84 & 1.25 & 1.83 & \\
$M$ in $10^5\,\msun$ & $0.34$ & $0.70$ & $1.66$ & \\
$r_0$ in $\rm pc$ & $1.0$&$1.0$&$1.0$&\\
$r_{\rm t}$ in $\rm pc$ & 6.91 & 18.0 & 68.1 & \\
$\sigma_0$ in $\kms$ & 5.0 & 5.0 & 5.0 & \\
$\rho_0$ in $10^3 \,\msun \rm{pc}^{-3} $&$ 4.1 $&$ 4.1  $&$ 4.1 $ &\\
\hline
\end{tabular}
\end{center}
\end{table}

\section{Modified Impulsive Approximation}
\label{mia}
\subsection{Impulsive Approximation with Point Particles}
To calculate the energy and momentum transfer between two point
particles due to solely gravitational interaction, it is
advisable to
study  the problem in the reduced co-ordinate system, i.e. one
separates the relative motion of the two particles from the center of
mass motion and follows the trajectory of the reduced particle as an
one-body problem (Landau~\&~\mbox{Lifschitz}~1976, Vol. I).
The net velocity change of  the reduced particle perpendicular and
parallel to the initial  direction  is:
\alpheqntext
\begin{eqnarray}
\left|\Delta \vv_{\perp}\right| \!\!&=&\!\! v_0 \sin
\theta = v_o |\sin 2\varphi_0|
\nonumber \\
\!\!&=&\!\! \frac{2 b v_0^{\,3}}{G(M + m)} \left[ 1 + \frac{
  b^2v_0^{\,4}}{G^2 (M + m)^2} \right]^{-1}
\end{eqnarray}
\begin{eqnarray}
  \left|\Delta \vv_{\parallel}\right| \!\!&=&\!\!  v_0 \left[ 1 - \cos
  \theta \right] = v_0 \left[ 1 + \cos 2\varphi_0 \right] \nonumber \\
  \!\!&=&\!\!  2v_0 \left[1 + \frac{ b^2v_0^{\,4}}{G^2 (M + m)^2}
\right]^{-1}
\end{eqnarray}
\reseteqntext

Here
$M$ and $m$ are the masses of the two particles and $\vartheta$ is the
scattering angle for the encounter. The variables $b$ and $v_0$
indicate impact parameter and initial relative velocity, respectively.
 $\Delta \vv_{\parallel}$ always
points in the  opposite direction of the initial velocity $\vv_0$ and $\Delta
\vv_{\perp}$ lies in the scattering plane.

The change of absolute velocity of particle $m$ is
\alpheqntext
\begin{eqnarray}
\label{1925a}
\left|\Delta \vv_{m{\perp}}\right|&=& \frac{M}{M+m}|\Delta
\vv_{\perp}| \nonumber \\
&=& \frac{2Mbv_0^{\,3}}{G(M+m)^2} \left [ 1 + \frac{
  b^2v_0^{\,4}}{G^2 (M + m)^2} \right]^{-1}\:,
\\
\label{1925b}
\left|\Delta \vv_{m{\parallel}}\right| &=& \frac{M}{M+m}\left|\Delta
\vv_{\parallel}\right| \nonumber \\
&=& \frac{2Mv_0}{M+m}  \left [ 1 + \frac{
  b^2v_0^{\,4}}{G^2 (M + m)^2} \right]^{-1}\:.
\end{eqnarray}
\reseteqntext

For the second particle $M$ one just has to interchange the
masses.

For large impact parameters, $b \rightarrow \infty$,  the
second term of the expression in brackets dominates and
equations~(\ref{1925a},b)
reduce to the {\em classical} impulsive approximation,
 which assumes a very short encounter time
such that one particle remains fixed in space whereas
 the other one moves along a straight line with constant velocity:
\alpheqntext
\begin{eqnarray}
\label{klassische-imp-approx}
\left|\Delta \vv^{\rm imp.}_{m{\perp}}\right| \!\!& =& \!\!\frac{2GM}{bv_0}\:;
\\
\left|\Delta \vv^{\rm imp.}_{m{\parallel}}\right|\!\!&=&\!\!  0\:.
\end{eqnarray}
\reseteqntext

Defining $\beta \equiv v_0^{\,2}b/G(M+m)$, these equations can be
written  as
\alpheqntext
\begin{eqnarray}
\label{imp-ubertrag-relativ}
\frac{\left|\Delta \vv_{\perp}\right|}{v_0}  \!\!& = &\!\! 2 \beta\:
\frac{1}{1+\beta^2}\:,\\
\frac{\left|\Delta \vv_{\parallel}\right|}{v_0}  \!\!&=&\!\! 2 \:
\frac{1}{1+\beta^2}
\end{eqnarray}
\reseteqntext
and
\begin{equation}
\label{imp-ubertrag-relativ-imp}
\frac{|\Delta \vv^{\rm imp.}_{\perp}|}{v_0}  = 2 \, \beta^{-1}\:.
\mbox{\hspace{1.0cm} (classical impulsive approximation)}
\end{equation}

\subsection{Influence of the Encounter Time}
\begin{figure}
\begin{minipage}[t]{8cm}
\mbox{\vspace{0.2cm}}
\end{minipage}
\begin{minipage}[t]{8cm}
\hspace{1cm}
\epsfxsize=8cm \epsfbox{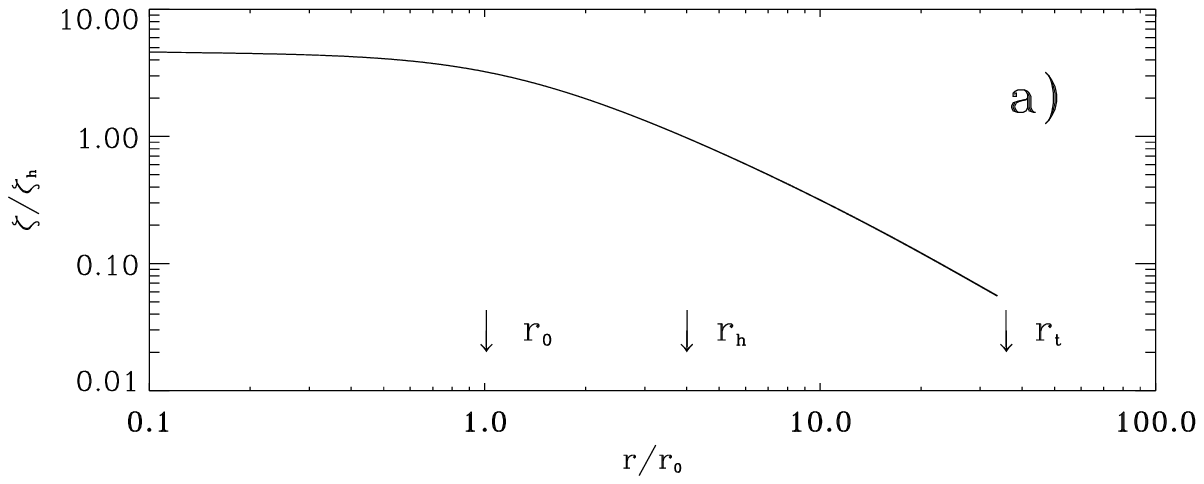}
\end{minipage}
\begin{minipage}[t]{8cm}
\mbox{\vspace{0.4cm}}
\end{minipage}
\begin{minipage}[t]{8cm}
\hspace{1cm}
\epsfxsize=8cm \epsfbox{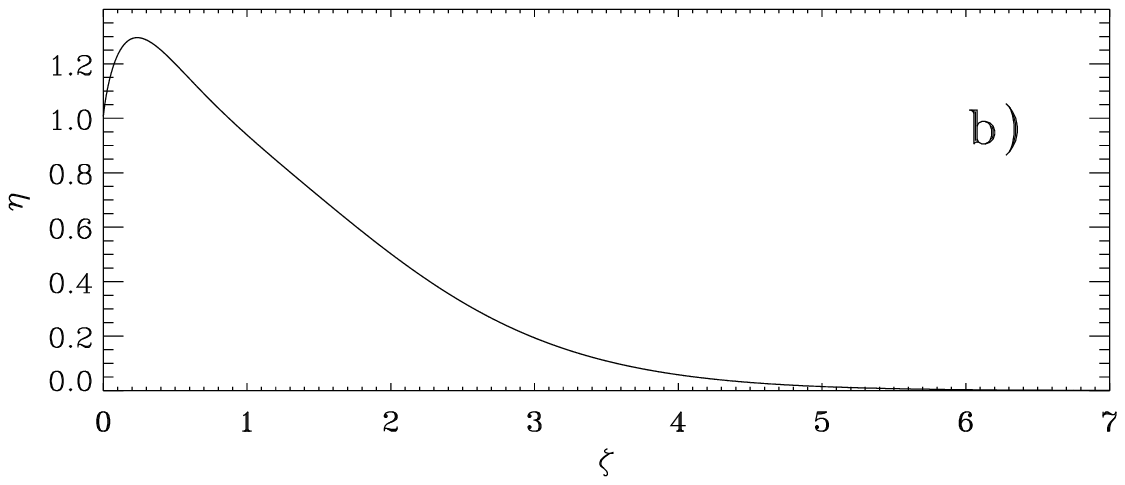}
\end{minipage}
\begin{minipage}[t]{8cm}
\mbox{\vspace{0.2cm}}
\end{minipage}
\caption[Relative Sto"szeit $\zeta$]{
\label{zeta-eta-pic}
a) The ratio $\zeta$ of effective interaction duration $\tau$ and
orbital period $t_{orb}$ as a function of radius, calculated for the
standard globular cluster and normalized to one at the half light
radius $r_{\rm h}$. b) Dependence of the correction factor $\eta$ on
$\zeta$.}
\end{figure}

For the (classical) impulsive approximation to be valid, the
encounter time $\tau$ has by far to be the shortest relevant time
scale in the system under study. The particle can then  be considered as
fixed in space and momentum is transfered quasi-instantaneously. Thus,
if a periodical oscillation of an individual star is superimposed
on the linear center of
mass trajectory of the whole system,
the orbital period $t_{orb}$ has to be much larger
than the time scale $\tau$ of the external perturbation ($\tau \ll
t_{orb}$).
In the adiabatical limit ($\tau \gg t_{orb}$), the orbital energy of
the star is not affected.
 The perturber ``sees'' the star as being smeared out over the whole
periodic trajectory and hence can only interact with its center of mass
averaged over the whole period; the orbit acts as a rigid body.

A very fast encounter of a massive black hole with a Galactic
globular cluster can therefore effectively alter both
the center of mass energy as well as
the internal energy (i.e. the velocity dispersion: $E_{int} \propto
\sigma^2$) of the cluster,  whereas
a very slow encounter  only affects the center of mass velocity
of the system.

The orbital periods of stars in globular clusters differ by many orders of
magnitude between stars in the core region and stars in the outer
envelope.  Thus for certain impact parameters $b$ and encounter
velocities $v_0$ the impulsive approximation will be valid in the outer
parts, whereas in the inner regions the adiabatic limit is reached.
A realistic application of the impulsive
approximation to  these objects has to take this into account.
Spitzer~(1958) defined the relative duration of the interaction $\zeta$ as
being
the fraction of the absolute encounter time $\tau$ relative to the
orbital period $t_{orb}$ of an individual
 cluster star: $\zeta \equiv \tau/t_{orb}$.
Under the assumption of a harmonic cluster potential he was able to
derive an analytical expression for a correction factor $\eta(\zeta)$
to parametrize the effectivity of the interaction ($\Delta {\bf v} /
{\bf v} \rightarrow \eta  \Delta {\bf v} /
{\bf v}$),
\begin{equation}
\label{def-spitzer-korrektur}
\eta \equiv \frac{1}{2}(L_x+L_y+L_z)\:,
\end{equation}
with
\alpheqntext
\begin{eqnarray}
L_x(\zeta) \!\!\!&=&\!\!\! \left[\zeta^2 K_1(\zeta) + \zeta K_0(\zeta)
\right]^2 +
\left[ \zeta^2 K_1(\zeta) \right]^2\:,\\
L_y(\zeta) \!\!\!&=&\!\!\! \left[\zeta^2 K_0(\zeta)\right]^2 +
\left[\zeta^2K_1(\zeta) \right]^2\:,\\
L_z(\zeta) \!\!\!&=&\!\!\! \left[\zeta K_1(\zeta) \right]^2\:.
\end{eqnarray}
\reseteqntext
Here the $K_n(\zeta)$ are the modified Bessel functions of second rank
and \mbox{n-th} order, the Macdonald functions. These follow for large
$\zeta$ the asymptotic formula \mbox{$K_n(\zeta)
= (\frac{1}{2}\pi \zeta^{-1})^{1/2} e^{-\zeta}$}{$\left[1+ \cal
O(\zeta^{-1})\right]$}
(Bronstein~\&~Semendjajew~1987).
So, for small encounter times $\tau \rightarrow 0$ (and thus $\zeta
\rightarrow 0$) the (classical) impulsive approximation is recovered and
for large durations ($\zeta \rightarrow \infty$) the adiabatical limit
is reached, which means
the effectivity of the interaction exponentially approaches zero.

Within the framework of this paper it is best suited to define the
encounter time $\tau$ as the time interval within which (under
idealized conditions, i.e. fixed particle location and straight moving
perturber)
$95\,\%$ of
the total momentum transfer takes place: $\tau \equiv \left.\Delta t_{
  }\,\right|_{95\%} \simeq 3 \, b/v_0$.
The stars of a globular cluster move -- under the assumption of circular
orbits -- with velocities $v_{\rm c}^{\,2} = r
d\Phi/dr = GM(r)/r$ in the potential $\Phi(r)$ where
$M(r) \equiv  4\pi\int_0^r \rho(r')r'^2dr'$
is the mass within the radius $r$.
The orbital period is then
\begin{equation}
\label{bahnperiode}
t_{orb} = \frac{2\pi r}{v_{\rm c}} = 2 \pi \sqrt{\frac{r^3}{GM(r)}}
\end{equation}
and the relative duration of the encounter follows as
\begin{equation}
\label{def-rel-zeit}
\zeta \equiv \frac{\tau}{t_{orb}} = \frac{3}{\pi}\frac{b}{v_0}\left(
\frac{GM(r)}{r^3}\right)^{1/2} \simeq \frac{b}{v_0}\left(
\frac{GM(r)}{r^3}\right)^{1/2} \:.
\end{equation}
Figure~\ref{zeta-eta-pic}.a shows the values of $\zeta$ for the
standard globular globular cluster (model \#1, see
Sect.~\ref{globulars-properties}) normalized to one at the half light
radius $r_{\rm h}$. The correction factor $\eta$
(equation~\ref{def-spitzer-korrektur}
 scales as
plotted in Fig.~\ref{zeta-eta-pic}.b. The slight increase of the
interaction efficiency for $\zeta \sil 0.8$ results from resonances
between orbit and perturber potential. For a large interaction
duration $\zeta \gg 1$ the adiabatical limit is reached, $\eta$
approached zero exponentially.

\begin{figure*}
\begin{minipage}[t]{16cm}
\epsfbox{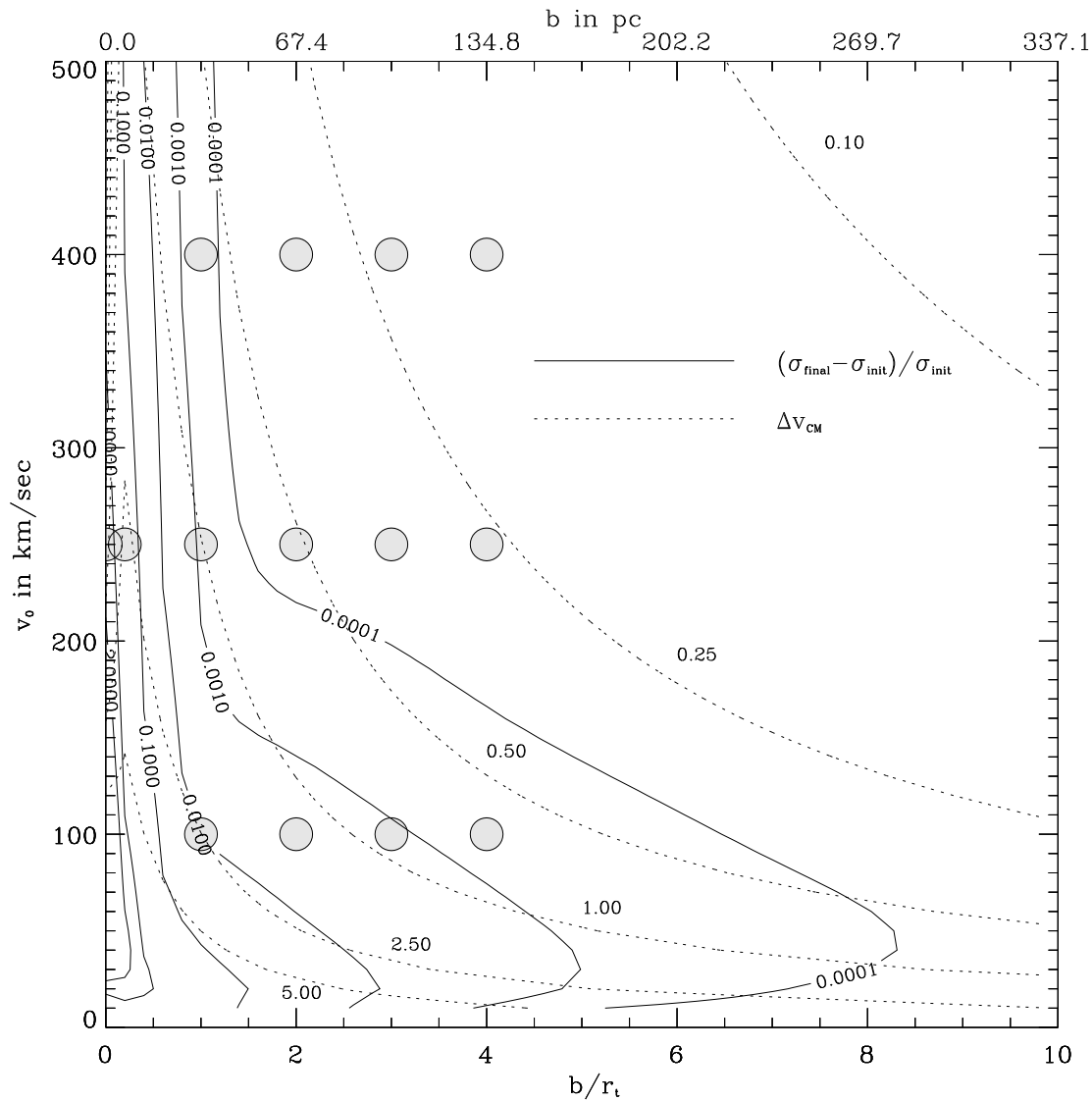}
\end{minipage}
\caption[]{
\label{imp-mit-1}
Relative increase of the velocity dispersion
$(\sigma_{final}-\sigma_{init})/\sigma_{init}$ (solid line) and the
center of mass velocity (dotted line; values in $\rm{km\,s}^{-1}$)
due to the encounter as function of impact parameter $b$ (normalized
to the tidal radius $r_{\rm t}$) and  impact velocity $v_0$ of the black
hole, computed using the modified impulsive approximation. The shaded
circles denote the parameter pairs, for which N-body simulations were
done. The total mass of the globular cluster is $1.03\times
10^5\,\rm{M}_{\odot}$ (model \#1: standard cluster with $r_{\rm t} = 33.71
\,\rm{pc}$) and the mass of the black hole is
$10^6\,\rm{M}_{\odot}$.
}
\end{figure*}

\begin{figure*}
\unitlength=1.0cm
\begin{picture}(16,19)
\put(0.3,0){\epsfxsize=15.2cm \epsfbox{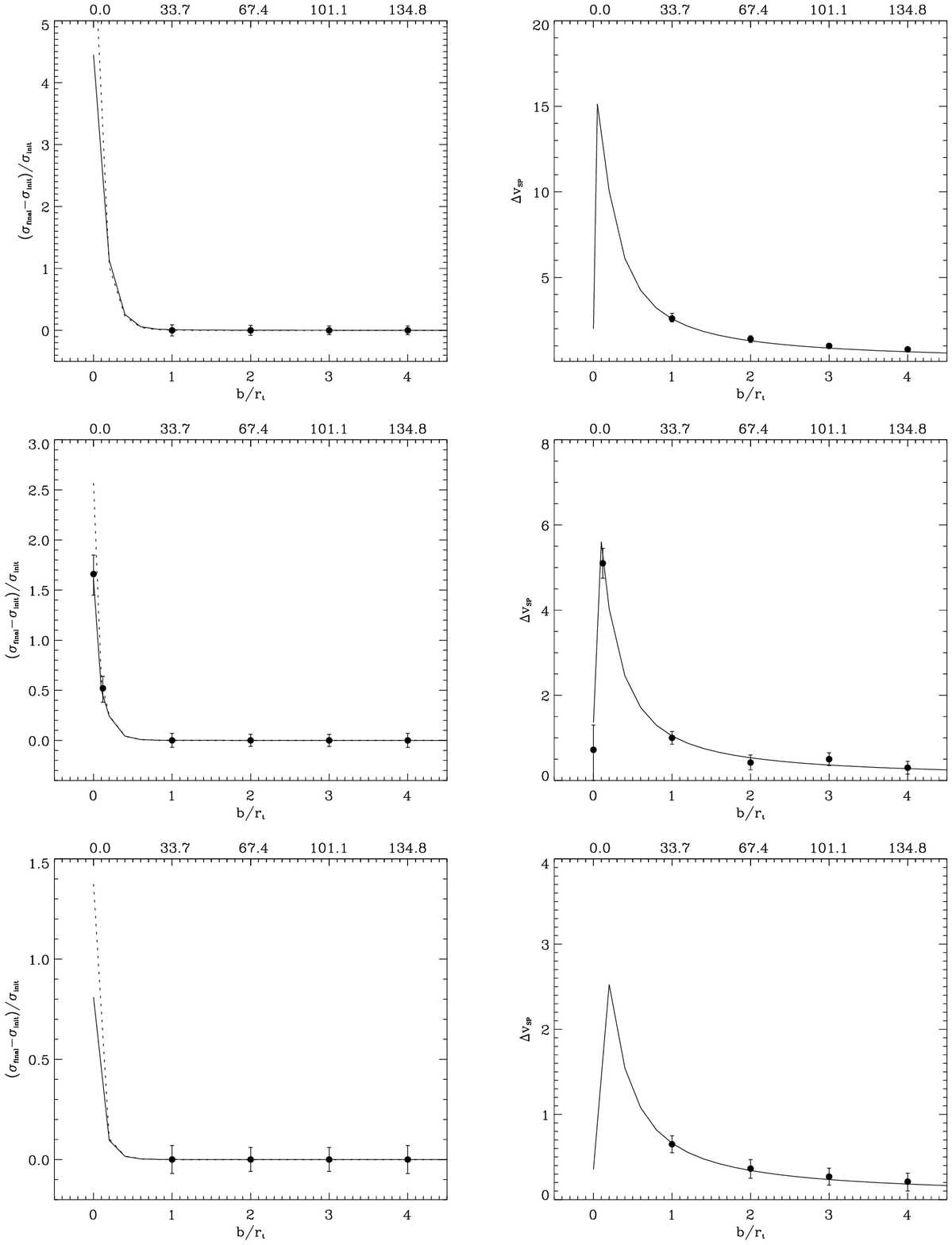}}
\put(3.8,4.7){\makebox{$v_0 = 400\,\rm{km\,s}^{-1}$}}
\put(11.8,4.7){\makebox{$v_0 = 400\,\rm{km\,s}^{-1}$}}
\put(3.8,11){\makebox{$v_0 = 250\,\rm{km\,s}^{-1}$}}
\put(11.8,11){\makebox{$v_0 = 250\,\rm{km\,s}^{-1}$}}
\put(3.8,17.4){\makebox{$v_0 = 100\,\rm{km\,s}^{-1}$}}
\put(11.8,17.4){\makebox{$v_0 = 100\,\rm{km\,s}^{-1}$}}
\end{picture}
\caption[]{
\label{vergl-imp-nbody}
Comparison of impulsive approximation and N-body-simulations for impact
velocities \mbox{$v_0 = 100\,\rm{km\,s}^{-1}$}, $v_0=
250\,\rm{km\,s}^{-1}$ and $v_0 = 400\,\rm{km\,s}^{-1}$.
The left column shows the relative increase of the velocity dispersion
and the right one the change of the center of mass velocity. The
result of the modified impulsive approximation is the solid line and the
points denote the N-body-simulations. For completeness, the dashed line
depicts the
uncorrected impulsive approximation.
}
\end{figure*}

\subsection{ Application to Globular Clusters -- Energy Transfer}
To calculate the total energy input by an encounter with a massive
black hole into a globular cluster as an extended object consisting of
many (~$10^5$) stars, one best applies the modified impulsive
approximation as derived above to every single star and then sums
up.  The increase of kinetic energy immediately after the
encounter is
\begin{equation}
  \Delta K^*_i = m_i \vv_i \cdot \Delta \vv_i + \frac{1}{2}m_i \Delta
  \vv_i^{\,2},
\end{equation}
where $\Delta K^*_i$ is the change of kinetic energy of star $i$ and
$m_i$, $\vv_i$ and $\Delta \vv_i$ are its mass, its  velocity
within the cluster prior to the encounter and the resulting velocity
change, respectively. The total change of kinetic energy $\Delta K$ is
obtained by the statistical average
\begin{equation}
  \Delta K = M \overline{\vv \cdot \Delta \vv} + \frac{1}{2} M
  \overline{\Delta \vv^2}\:.
\end{equation}
$\Delta K$
contains the center of mass acceleration of the whole cluster as well
as its increase of internal energy. The later is the
parameter of interest in the present context. It is to be compared with
the binding energy of the system $E_B$.

Since in this approach the stars within the cluster are considered as
fixed in space during the encounter, the potential energy is not altered.
Due to the increase of kinetic energy,
the globular cluster has lost its virial equilibrium and
begins to relax towards a new equilibrium state.  Throughout this
process kinetic energy is converted into potential energy and the
cluster as a whole expands.and vice versa.
Thereby it eventually
evaporates stars through its tidal barrier and in strong interactions
($\Delta K \sig E_B$) dissolves completely
(Chernoff~\&~Weinberg~1990).

\subsection{Comparision with Numerical N-Body Simulations}
\label{umpf-1}
\label{comp}
To test the accuracy of the above  approximation, we have
calculated the encounters of the standard globular cluster (model \#1, see
Sect.~\ref{globulars-properties}) with a black hole of mass
$\mbh=10^6\,\msun$ for 14 different pairs of impact parameter $b$ and
relative velocities $v_0$ using a standard {\sc Tree~Code} scheme
 (e.g. Barnes~and~Hut~1986,  Hernquist~1987,~1990). We
compare its result with the parameter study undertaken with the
modified impulsive approximation described above.
Figure~\ref{imp-mit-1} shows the relative increase of the internal
velocity dispersion $(\sigma_{final}-\sigma_{init})/\sigma_{init}$ and
of the center-of-mass velocity $\Delta v_{\rm CM}$ of the cluster in
$\rm{km\,s^{-1}}$ due to
the encounter, as predicted by the impulsive approximation.
 The encounter speed varies from
$10\,\rm{km\,s^{-1}}$ to $500\,\rm{km\,s^{-1}}$, the impact
parameter -- scaled in units of the tidal radius of the globular
cluster -- from $0$ (black holes penetrates through the center of
the cluster) via 1 (the black hole just touches the cluster) to 10
(distant encounter). The adiabatical correction factor has a strong
influence for velocities smaller than $250\,\rm{km\,s}^{-1}$. The
amplification of the  encounter effectiveness $\eta$ is due to
resonances between perturber and the orbits of cluster stars (see
equation~(\ref{def-spitzer-korrektur}) or Fig.~\ref{zeta-eta-pic}) and
pushes the lines of equal increase of velocity dispersion to larger
impact parameters $b$.
In the upper half of the figure. i.e. for large encounter speeds,
the adiabatical correction is negligible and the original hyperbolic
shape of the lines $\Delta \sigma / \sigma = {\rm const.}$ is
recovered.
The center of mass motion is not affected by internal adiabatic
effects. Therefore the lines of constant $\Delta v_{\rm CM}$ show the
expected hyperbolic behavior throughout the whole considered parameter range.
 The shaded circles denote  the parameter pairs
for which additional N-body simulations were made.
The results of these simulations were compared with the
analytical predictions in Fig.~\ref{vergl-imp-nbody}. The numerically
derived values for $(\sigma_{final}-\sigma_{init})/\sigma_{init}$ and
$\Delta v_{\rm CM}$ agree remarkably well with the analytically
derived ones. Therefore the modified impulsive
approximation is an adequate tool to determine the increase of the internal
energy of a globular cluster due to such encounters and thus the
probability for the cluster to get disrupted.

\section{Disruption Time Scales for the Standard
  Cluster}
\label{discussion-I}
We adopt in our calculations a mass model for the Milky Way, which
contains a bulge and disk component and a dark isothermal halo to
account for the
flat rotation curve with $v_{rot} \simeq 220 \: \rm{km\,s}^{-1}$.
The bulge is analytically represented by a Plummer spheroid with total
mass $M_B = 1.5 \times 10^{10}\:\rm{M}_{\odot}$ and core radius
$r_B = 1.5 \: \rm{kpc}$. The disk component is calculated as
double exponential with radial scale length $R_D = 3.5 \:
\rm{kpc}$ and vertical scale height $z_D = 0.25 \: \rm{kpc}$.
Its density distribution is normalized to $\rho_{\odot} = 0.072
\:\rm{M}_{\odot}\rm{pc}^{-3}$ in the solar neighborhood (Allen
1973,  \S 118). The dark halo results then in
\begin{equation}
\label{hallali}
\rho(r) = 0.134\,\frac{\rm{M}_{\odot}}{\rm{pc}^3}\,\left(
1+\frac{r^2}{R_{\rm H}^{\,2}} \right)^{-1}\:,
\end{equation}
where $R_{\rm H}=2.5\,\rm{kpc}$ denotes the core radius. In the solar
vicinity, i.e. for $r=R_{\odot}=8.5\,\rm{kpc}$, one obtains a
dark matter density of $\rho(R_{\odot}) =
0.01\:\rm{M}_{\odot}\,\rm{pc}^{-3}$.  Following the
considerations of Lacey~\&~Ostriker~(1985) that this dark halo
consists entirely of massive black holes with $\mbh
=10^6\,\rm{M}_{\odot}$, there are about 10 of these objects in a
volume of $1\,\rm{kpc}^3$ around the sun. In the center of the
Galaxy we would expect approximately $134$ black holes
per $\rm{kpc}^3$. The $1/r^2$- density fall-off resembles
that of an isothermal sphere and thus one can derive a velocity
dispersion for these objects of
\begin{equation}
\label{hallala}
\sigma_{\rm BH} = \sqrt{2\pi G r^2 \,\rho(r)} \simeq 120 \rm{km\,s}^{-1}\:.
\end{equation}
The velocity distribution   of the black holes in one direction is
then assumed to be
Gaussian with zero mean and a standard deviation of $120 \,\rm{km\,s}^{-1}$.

We assume, that the globular clusters move on circular
orbits with $v = v_{\rm c} \simeq 220\:\rm{km\,s^{-1}}$ through this background
medium of black holes.
This is a
somehow simplistic ansatz, but adequate for the so called thick disk
population of globular clusters and a reasonable order of magnitude
approximation for the old halo clusters.
Then the number of encounters between these
black holes and an individual globular cluster at a galactic radius
$R$  with impact parameters
in the interval $[b,b+db]$ and impact velocities $[v,v+dv]$ within
a period $t$ is
\begin{eqnarray}
\lefteqn{N(R;b,v,t)dbdv =}  \nonumber\\
&=&\!\!\!2\pi
bdb\cdot\left[\frac{v_{\rm c}}{\sqrt{2\pi\sigma_{\rm BH} ^2}}\,
e^{\textstyle -\frac{1}{2}\frac{(v-v_{\rm c}) ^2}{\sigma_{\rm BH}^2}}\right]dv
\cdot t \cdot \frac{\rho(R)}{M_{\rm BH}}\:.
\end{eqnarray}
The expression in brackets accounts for the Gaussian distribution of
black hole velocities.
Thus in the Milky Way the total number of encounters with $b \le \tilde{b}$
within
the time interval $t$ is
\begin{eqnarray}
\lefteqn{N(R;b\le \tilde{b} ,t)=} \nonumber \\
&=&\!\!\!\int\limits_0^{\tilde{b}}db\int\limits_{-\infty}^{+\infty}dv\,
N(R;b,v,t)\nonumber\\
&=&\!\!\!\pi \tilde{b}^2\cdot v_{\rm c}\cdot t \cdot
\frac{\rho(R)}{M_{\rm BH}} \nonumber \\
&=&\!\!\!0.947\left(1+\frac{R^2}{R_{\rm H}^{\,2}}\!
\right)^{-1}\left[\frac{M_{\rm BH}}{10^6\,\rm{M}_{\odot}}\right]^{-1}
\left[\frac{\tilde{b}}{\rm{pc}^3}
\right]^2\left[\frac{t}{10^{10}\rm{a}}\right]\:.
\end{eqnarray}
The resulting numbers of encounters $N$ between black holes with
$M_{\rm BH}=10^6\,\rm{M}_{\odot}$ and a standard cluster (model
\#1,
see
Tab.~\ref{gc-models}) at galactocentric
distances $R= 5\,\rm{kpc}$,
$R= 10\,\rm{kpc}$
and $R= 15\,\rm{kpc}$ within a period of $t = 10^{10}$ years are given
in Tab.~\ref{table-n}.

\begin{table}
\caption{
\label{table-n}
Average Number $N$ of encounters between the standard cluster
(model \#1) on
circular orbits at $R= 5\,\rm{kpc}$,
$R= 10\,\rm{kpc}$
and $R= 15\,\rm{kpc}$ and
black holes with masses $M_{\rm BH}=10^6\,\rm{M}_{\odot}$ during
$t=10^{10}$ years depending on the impact parameter $\tilde{b} \equiv
b/r_t$, normalized to the tidal radius $r_{\rm t}$ of the cluster
($r_{\rm h}$ denoting the half mass radius).
}
\begin{center}
\tabcolsep0.15cm
\renewcommand{\arraystretch}{1.5}
\begin{tabular}[t]{|rcl|c|c|c|}
\hline
\multicolumn{3}{|c|}{$\tilde{b}$} & $N(5\,\rm{kpc})$ & $N(10\,\rm{kpc})$ &
 $N(15\,\rm{kpc})$ \\ \hline \hline
\renewcommand{\arraystretch}{1.1}
$r_{\rm h}$ & = & $4\,\rm{pc}$ & $3$ & $1$ & $0.4$ \\
$\frac{1}{2}\,r_{\rm t}$ & = & $16.8\,\rm{pc}$ & $54$ & $16$ & $7$ \\
$1\,r_{\rm t}$ & = & $34\,\rm{pc}$ & $215$ & $63$ & $29$ \\
$2\,r_{\rm t}$ & = & $67\,\rm{pc}$ & $860$ & $253$ &$116$ \\
$5\,r_{\rm t}$ & = & $160\,\rm{pc}$& $5\,377$ & $1582$ & $727$
\\
\hline
\end{tabular}
\end{center}
\end{table}

During its lifetime a globular cluster would experience very many
encounters, even in the outer parts of the Galaxy. Clearly this
bombardment would have devastating effects on the population of
Galactic globulars, as will be quantified below.

To compute the gain of kinetic energy due to interactions with
black holes, one has to convolve the number of encounters $N(R;b,v,t)$
with the increase of the internal velocity dispersion per encounter
$\left[(\sigma_{final}-\sigma_{init})/\sigma_{init}\right](b,v)$ of
the globular cluster:
\begin{eqnarray}
\label{miese-ratte}
\lefteqn{\left[\frac{\sigma_{final}-\sigma_{init}}{\sigma_{init}}\right](R;
t)=} \nonumber \\
&=&\!\!\int\limits_0^{\infty}db\int\limits_{-\infty}^{\infty}dv\,N(R;b,v,t)
\cdot \left[\frac{\sigma_{final}-\sigma_{init}}{\sigma_{init}}\right]
(b,v) \nonumber\\
&=&\!\!\int\limits_0^{\infty}db \int\limits_{-\infty}^{\infty}dv\,2\pi b
\cdot \left[\frac{v_{\rm c}}{\sqrt{2\pi\sigma_{\rm BH}^2}}\,e^{\textstyle
  -\frac{1}{2}\frac{(v-v_{\rm c})^2}{\sigma_{\rm BH}^2}}\right]
\nonumber \\
&& \:\:\:\:\: \:\:\:\:\: \:\:\:\:\: \:\:\:\:\: \:\:\: \times \:\:t \cdot
\frac{\rho(R)}{M_{\rm BH}}\cdot
\left[\frac{\sigma_{final}-\sigma_{init}}{\sigma_{init}}\right](b,v)
\end{eqnarray}

Using the values of the
standard cluster (model \#1 with $[\Delta \sigma / \sigma ](b,v)$ according to
Fig.~\ref{imp-mit-1} and $\mbh =10^6\,\msun$), this results in
\begin{eqnarray}
\label{miese-latte}
\lefteqn{\left[\frac{\sigma_{final}-\sigma_{init}}{\sigma_{init}}\right](R;
t)\:=} \\
&=&\!\!\! 80.6\left(1+\frac{R^2}{R_H^{\,2}}\right)^{-1}
\left[\frac{M_{\rm BH}}{10^6 \,\rm{M}_{\odot}} \right]^{-1}
\left[\frac{r_{\rm t}}{33.7\,\rm{pc}}\right]^2
\left[\frac{t}{10^{10}\,\rm{a}}\right] \nonumber
\end{eqnarray}

\begin{figure}
\unitlength1.0cm
\epsfxsize=8cm \epsfbox{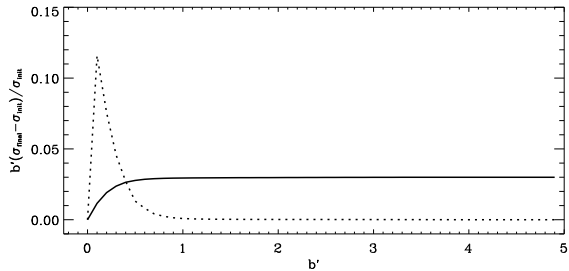}
\caption[]{
\label{faltung}
Function  $b' \Delta \sigma / \sigma$
(dotted line) and the resulting integral (solid line) depending on the
relative impact parameter $b'=b/r_{\rm t}$ at fixed $v=v_{\rm c}$ for the
standard
model of a globular cluster in the Milky Way.}
\end{figure}

How well does the integral in equation~\ref{miese-ratte} converge and which
impact parameter dominate the result? These questions are addressed in
Fig.~\ref{faltung}.  It
shows the integrand $b'
\left[\Delta \sigma/\sigma\right]$
(dotted line) at fixed $v=v_{\rm c}$ and the resulting integral
(solid
line), as function of  $b'=b/r_{\rm t}$, the impact parameter normalized to the
cluster radius.  Saturation and thus convergence of the integral
is  reached already within the cluster itself. This indicates
that only a relatively small number of  central encounters with
$b' < 1$ dominates the total energy gain.

The above  calculation shows a large increase of the velocity dispersion of
the standard cluster over the lifetime of the Milky Way, e.g. by a
factor of 4.7 at a galactocentrical distance $R
= 10 \:\rm{kpc}$.  No cluster would survive such a
bombardment. But what is the scatter around the mean of this value?
Equation~(\ref{miese-ratte}) was obtained by integrating over a
{\em smoothed} background distribution of dark halo objects. Under the
assumption of black holes being the main constituent of the dark halo,
this smoothness is no longer true: the distribution of matter in the
halo is expected to be extremely coarse grained.  Furthermore, the
evolution of the
globular cluster is mainly affected by encounters within the innermost
region and these are relatively rare (as shown above).
Which means the collision history and the resulting dispersion gain is
strongly determined by only a handful events and thus one expects huge
deviations from the mean for different encounter histories.
Therefore, even if the average energy gain would rule out the survival
of a certain type of globular clusters within reasonable distances from
the galactic center, individual clusters might still be able to resist
this destruction mechanism due to the expected large scatter in
$\Delta \sigma / \sigma$ as a result of different encounter histories.
 A realistic study of the survival rate
within a population of specific  globular clusters must take this effect into
account and apply statistical methods.
This shall be done in the next section, which describes  the Monte Carlo
simulation of individual cluster histories within the Galaxy.

\begin{figure*}
\begin{minipage}[t]{16cm}
\epsfxsize=16cm \epsfbox{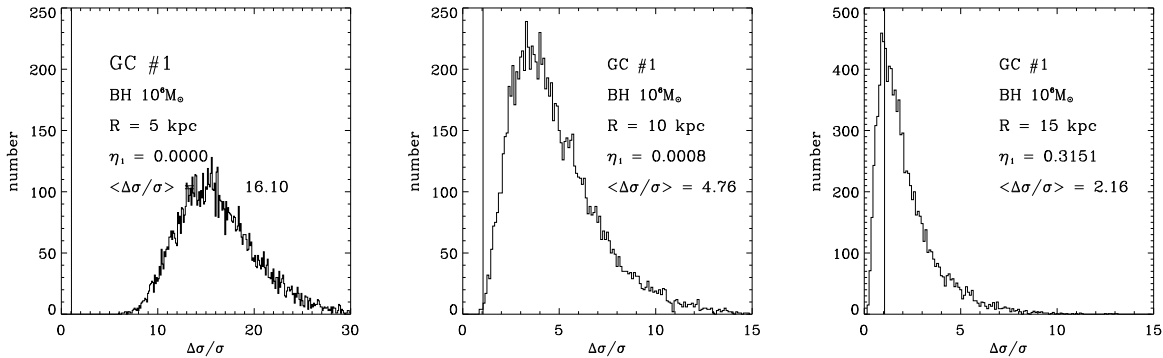}
\end{minipage}
\caption[]{
\label{wilder-aff}
Histogram of the dispersion gain $\Delta \sigma / \sigma$ of the
standard globular cluster (model \#1) for a black hole mass of
$10^6\,\msun$ at $R= 5$, $10$ and $15 \,{\rm kpc}$ ($10\,000$
encounter histories per plot). Histories that lead to $\Delta \sigma
/ \sigma < 1$ are counted as cluster survivals. $\eta_1$ denotes the
fraction of these events. The average gain $\langle \Delta \sigma /
\sigma \rangle$ agrees well with the analytical value obtained in
equation~(\ref{miese-latte}) at these radii.
}
\end{figure*}

\section{Monte Carlo Simulations of Individual Encounter Histories}
The mean free path of particles with velocities  $\langle v\rangle$
in a sea of
background particles with number density $n$ and cross
section $\Sigma$ is $\langle l \rangle = (n \Sigma)^{-1}
= \langle v\rangle \tau$.  Thus the average time $\tau$
between two encounters with given impact parameter in the interval
$[b,b+\Delta b]$ and relative velocity $v$ is
\begin{equation}
\label{hallalu}
\tau = \frac{1}{n \Sigma v}\:.
\end{equation}
The number density of halo black
holes is $n(R) = \rho(R) / M_{\rm BH}$, adopting a halo density
distribution $\rho$, according
to equation~(\ref{hallali}). The interaction
cross section is $\Sigma = 2 \pi b \Delta b$ for simple geometric
arguments. Since the distribution of dark halo objects is assumed to
be isothermal with velocity dispersion $\sigma_{\rm BH} \simeq
120\:\kms$ (see equation~\ref{hallala}), the distribution of encounter
velocities between the halo objects and a globular cluster
is a Gaussian with standard deviation $\sigma_{\rm BH}$
around the center of mass velocity of the globular cluster.
 Assuming they  are moving
on circular orbits with $v_{\rm c} \simeq 220\:\kms$, one has to weight the
velocity entering equation~(\ref{hallalu}) with $(2\pi
\sigma_{\rm BH}^2)^{-1/2} \exp(-\frac{1}{2}(v-v_{\rm c})^2/\sigma_{\rm
BH}^2)$.
This way we obtained the typical time interval $\tau(b,v)$ between two
encounters for the whole parameter region described in
Sect.~\ref{comp}.

Transport theory (consult e.g. Landau~\&~Lifschitz~1983, Vol. X) specifies the
probability for a particle {\em not} to undergo an encounter in the
time interval $\Delta t$ as $\tilde{P}(\Delta t) = e^{-\Delta t/\tau}$.
Therefore,
the probability that a
globular cluster
actually {\em does} encounter a black hole within $\Delta t$ and with
interaction
parameters $(b,v)$, results as $P(\Delta t) = 1- \tilde{P}(\Delta t) = 1 -
e^{-\Delta t/\tau(b,v)}$.
The time interval $\Delta t$ corresponding to  a given probability $P$
 is then $\Delta t(P) = - \tau \ln (1- P)$.
Since encounters with different impact parameters $b$ and $v$ are
statistically independent, one can handle each grid point $(b,v)$ in
parameter space separately. For each  pair $(b,v)$ we throw a dice  to
obtain a random probability $P$ and translate this random number into a
time interval $\Delta t(P)$ as described above. If $\Delta t(P)$ is smaller
than
the studied total time interval $t_0$ (here initially $10^{10}$ years),
then we accept the
encounter event, i.e.  add $\Delta \sigma(b,v)$ to the velocity
dispersion of the cluster and substract $\Delta t(P)$ from $t_0$, leading to
a remaining time $t = t_0 - \Delta t(P)$.
This procedure is repeated, until the time interval $\Delta t(P)$ to the next
encounter
becomes larger than the remaining time $t$.

Figure~\ref{wilder-aff} visualizes the distribution of $\Delta
\sigma / \sigma$ after $10^{10}$ years for the standard cluster (model
\#1) in a galactic dark halo of $10^6\:\msun$ black holes at three
different galactic radii. The average dispersion gain  $\Delta \sigma /
\sigma$ at these radii agrees well with the value obtained in the last
section, using a smooth background distribution (equation~(\ref{miese-latte}).
 However, the
histogram shows the expected huge scatter due to the discreteness of
the encounter events. Adopting the binding energy of the cluster as
threshold for its disruption, we determine the fraction $\eta_1$ of
the clusters that get destroyed in the simulation: Whenever the total
kinetic energy of the cluster stars exceeds the binding energy of the
system, i.e. when $\Delta \sigma / \sigma > 1$, we assume the cluster
has dissolved over the studied time interval.  This threshold of
$\Delta \sigma / \sigma > 1$ is a somehow crude approximation.
We therefore have performed
 N-body calculations of encounters with $\Delta \sigma / \sigma
\sig 1$, which indicate that the mass loss during such an event is so
severe, that the cluster either dissolves instantaneously or the
remaining (bound) stars relax back into an equilibrium state
that is much less viable and
will get disrupted within a few $10^8$ years.
Clusters that suffered a central encounter with $\Delta \sigma / \sigma >
1.5$ were always destroyed.
An event with $\Delta \sigma / \sigma = 1.25$ typically causes the
cluster to loose $25\:\%$ of its stars. The energy input in a
central encounter is not  distributed evenly onto the cluster stars, but
affects those most that are located along the trajectory of the black
hole through the cluster. $75\:\%$ of the stars remain bound and
 form a new globular cluster like configuration within a few
relaxation time scales $\tau_{\rm rel}$. The binding energy of this
 ``new'' cluster however is much smaller than that of the original one by
a factor of five. Therefore the next encounter of similar strength (or
several more distant encounters with the same cumulative energy
input) will easily disrupt the cluster. After all, it is save to
take  the value $\Delta \sigma /\sigma = 1$ as an adequate upper limit
to cluster survival.

The simulations also indicate that a series of
distant (and therefore weak) encounters integrated over a certain
period $\Delta t$
 have a more serious effect on a globular cluster than
one single central encounter at the beginning of this time interval
$\Delta t$ leading to the same net energy input.
In the first case, the excess energy is more or less evenly
distributed onto the cluster stars thus leading to a high evaporation
rate throughout the whole studied time interval. Whereas in the case
of a central collision, some stars (alongside the trajectory of the black
hole through the cluster) gain very high velocities
and immediately leave the cluster taking away a large fraction  of the
energy input. The remaining stars stay kinematically colder and can  --
when the impact was not completely devastating in the first place
-- relax back into equilibrium, then again exhibiting  the small mass loss
rate of  unperturbed dynamical evolution.

The exemplary case, standard cluster at $R=10\,{\rm kpc}$
with black holes of $M_{\rm BH} = 10^6\,\msun$, reveals a extremely
low rate of survivors within the inner $10 \,{\rm kpc}$ of the Milky
Way (Fig.~\ref{wilder-aff}).
This analysis is extended to all models and to different black
hole masses in the next section.

\begin{figure*}
\begin{minipage}[t]{16cm}
\epsfxsize=15.5cm \epsfbox{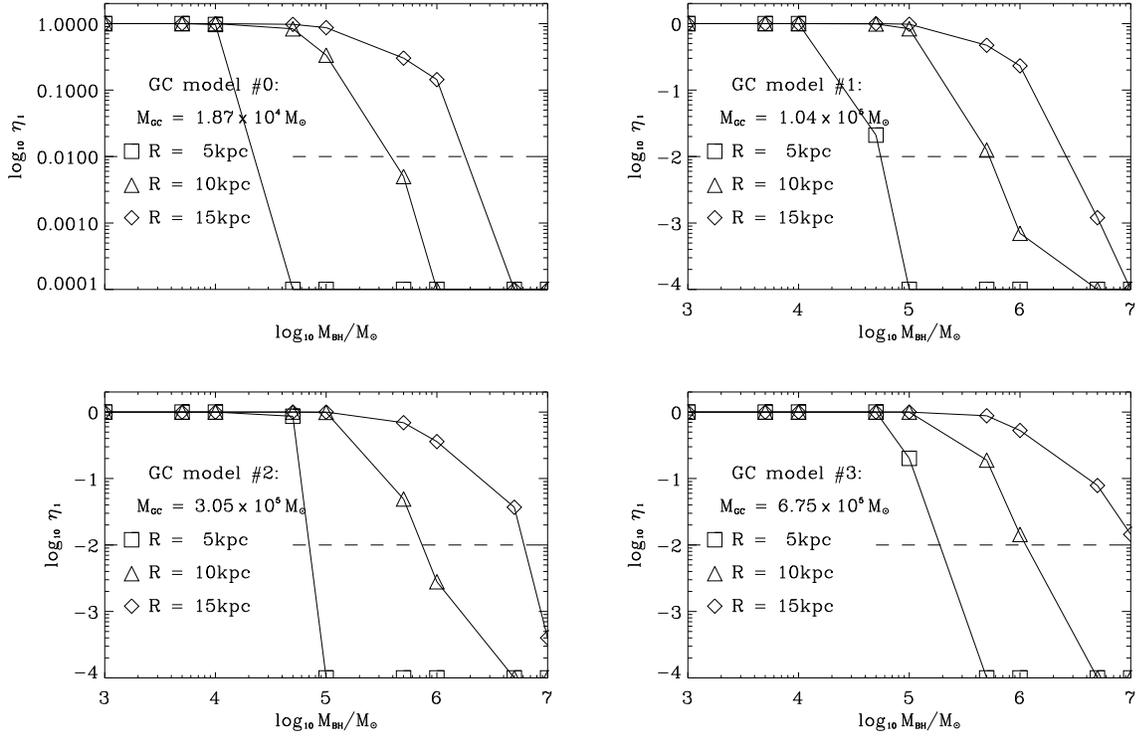}
\end{minipage}
\caption[]{
\label{plot-lifenum-mass-all-I}
 Survival rates $\eta_1$ for the model sequence \#0 to \#3, all with
 $c=1.53$, at three galactocentric distances.
}
\end{figure*}
\begin{figure*}
\begin{minipage}[t]{16cm}
\epsfxsize=15.5cm \epsfbox{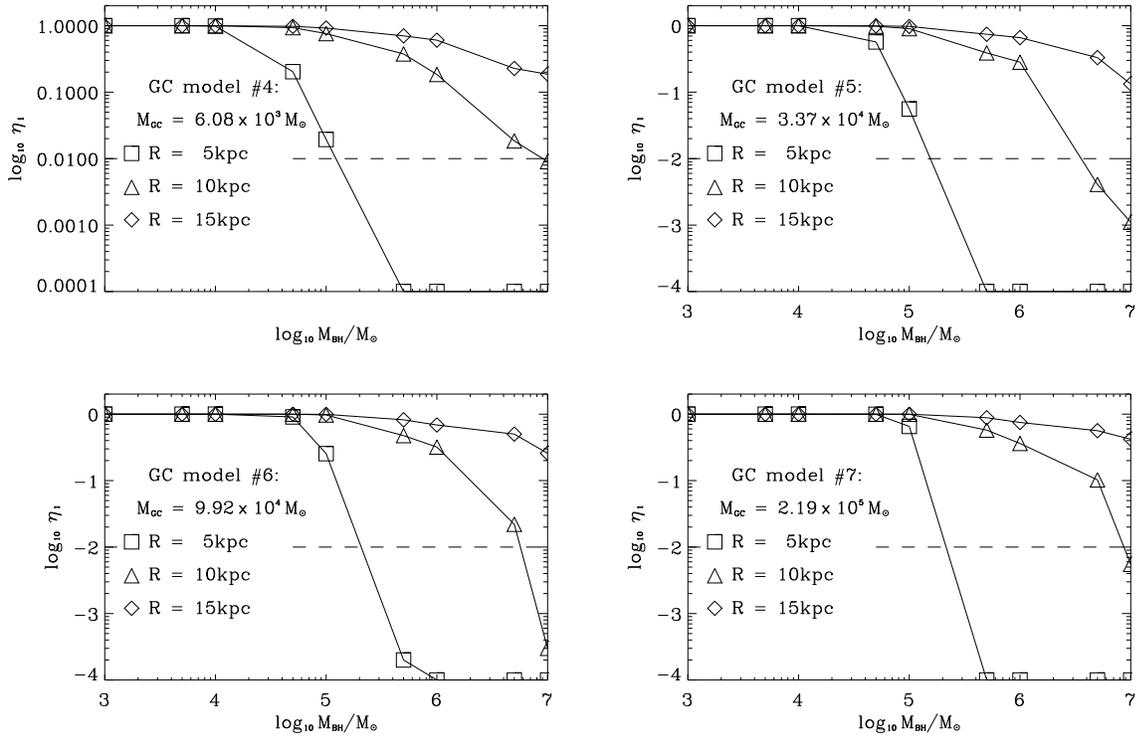}
\end{minipage}
\caption[]{
\label{plot-lifenum-mass-all-II}
The same, but for the  sequence \#4 to \#7, all with  $c=0.84$.
}
\end{figure*}
\begin{figure*}
\begin{minipage}[t]{16cm}
\epsfxsize=15.5cm \epsfbox{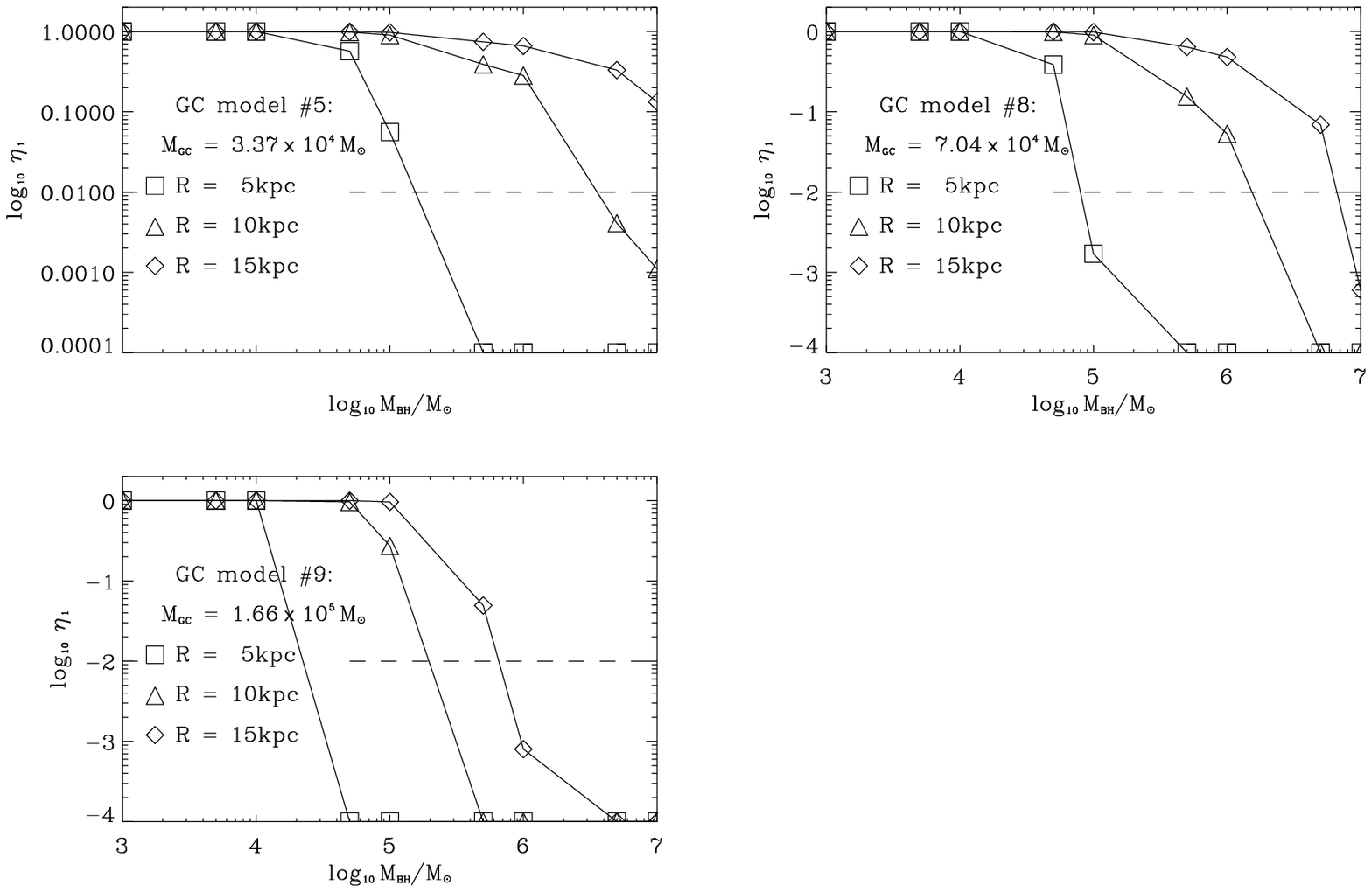}
\end{minipage}
\caption[]{
\label{plot-lifenum-mass-all-III}
Survival rates for the globular cluster models \#5, \#8 and \#9 (the
central velocity dispersion $\sigma_0$ is fixed to $5\,\kms$).
}
\end{figure*}

\section{Encounter Histories for Different Globular
  Cluster Models in the Milky Way}
\label{discussion-II}
Determinations of the total mass in the luminous halo of our Galaxy
lead to a value around $1.5 \times 10^9\,\msun$ (Carney~et.~al.~1991;
or e.g.  $M_{\rm H} = 9 \times 10^8\,\msun$ in the range $4$ --
$25\,{\rm kpc}$ in Suntzeff~et.~al.~1990). With a total mass of
around $2\times 10^7 \,\msun$ in (halo) globular clusters out to $R =
40\,{\rm kpc}$ (Carney~et.~al.~1991), the cluster
to halo mass ratio is 0.015 -- 0.02.  The maximum allowed cluster
disruption rate is obtained, assuming
 that all halo field stars are  the debris of
dissolved Galactic globular clusters, implying that the Galactic globular
cluster system was about a factor of 100 more massive when it formed more
then $10^{10}$ years ago than it is today. Neglecting the evolutionary
differences of  globular clusters of different masses and
concentrations, the above ratio sets a very conservative lower limit to
the survival rate of these clusters, which further can be used to
obtain an upper limit on the mass of dark halo black holes: If the
assumption of a certain black hole mass leads to a survival rate below
$\eta_1 = 0.01$, then this mass has to  be discarded.  With the method
described in the last section we calculate for each of the ten
globular cluster models the fraction of survivors for black hole
masses in the range $10^3\;\msun$ to $10^7\,\msun$ at three
galactocentric radii, $R=5\,{\rm kpc}$, $10\,{\rm kpc}$ and $15\,{\rm
  kpc}$. The result of these computations is plotted in
Fig.'s~\ref{plot-lifenum-mass-all-I},
\ref{plot-lifenum-mass-all-II},\ref{plot-lifenum-mass-all-III}. The
dashed line   indicates the acceptance threshold at $\eta_1 = 0.01$.

\begin{figure*}
\begin{center}
\begin{minipage}[l]{17cm}
\epsfxsize=15.5cm \epsfbox{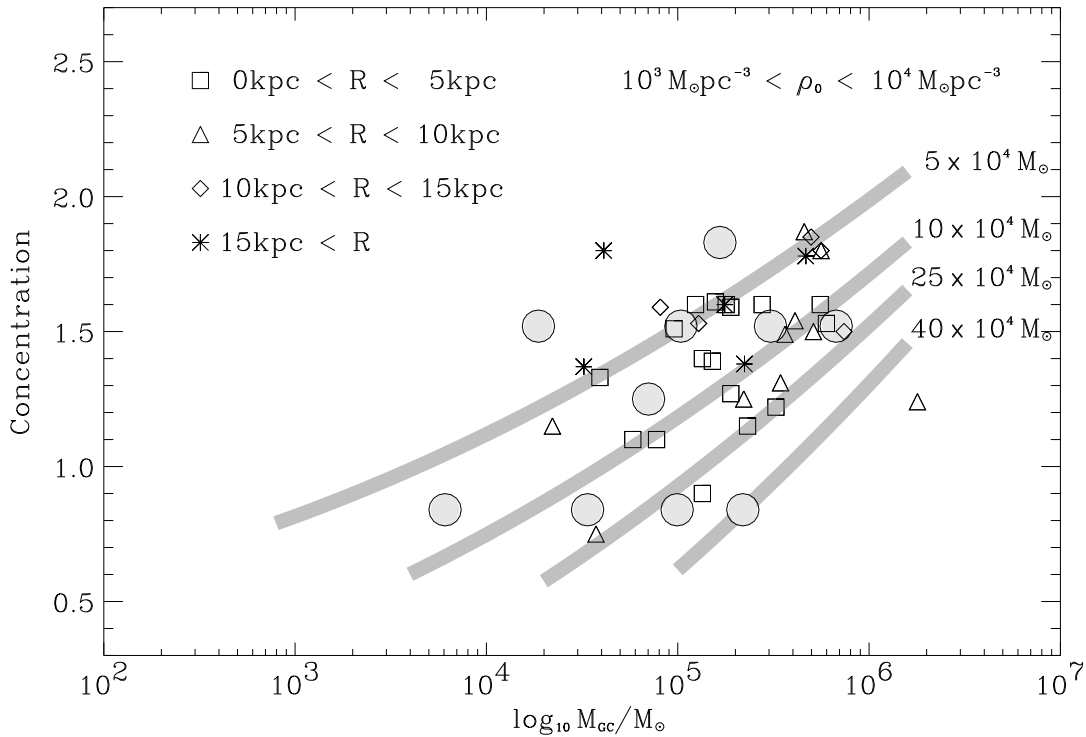}
\end{minipage}
\end{center}
\caption[]{
\label{final-limit}
Distribution of mass $M_{\rm GC}$ and concentration $c$ for the ten
model clusters and for Galactic globular clusters with central
densities in the range $10^3\:\msun \rm{pc}^{-3} < \rho_0 <
10^4\:\msun \rm{pc}^{-3}$ (Webbink~1985). The probability for globular
clusters to
survive is less then 1\%, i.e. $\eta_1 < 0.01$, in the region left and
above the gray lines for dark matter halos consisting of massive black
holes of $M_{\rm BH} = 5\:10^4\,\msun$, $10^5\,\msun$,
$2.5 \times 10^5\,\msun$ and $4 \times 10^5\,\msun$,
respectively. These lines were computed at $R = 5 \rm{kpc}$ and have
to be compared with the observed globulars with the same
galactocentric distance (squares).
}
\end{figure*}

These figures exhibit the following trends in our model:
\begin{enumerate}
\item
  Closer to the Galactic center the number density of dark halo
  objects increases and so does the number of encounters. Thus the
  range where the number of surviving globular clusters drops
  to zero shifts to lower black hole masses.
\item
  Within one sequence of equal concentration, the more massive
  clusters are more viable.
  This is understandable, because the bigger the cluster is, the
  tighter it is bound at fixed concentration (i.e. the deeper is its
  own potential well) and the better can it resist the steady
  bombardment of background particles.
\item
  Since $M \propto \rho_0$ and also $M \propto \sigma^2$ (according to
  equation~\ref{umpf_umpf}), the stability of the cluster increases with
  growing central density or dispersion.
\item
  Quite unexpectedly, we find that the most stringent
  constraints come from the most centrally
  concentrated clusters \#9 and \#0 to \#3.
  The larger $c$, the more extended is the envelope of less strongly
  bound stars outside the core and thus the geometrical cross section of
  the cluster. With an extended outer region a cluster can more effectively
  loose stars than a less centrally concentrated and thus smaller
  one. Variations
  of the concentration index $c$ have the largest effect
  on the survival rate of the model clusters. It even outweighs the
  influence of the cluster mass, as can be seen in sequence \#4, \#8 and
  \#9: The threshold of 1\% survivors is reached for cluster \#9 first,
  even though it has the highest mass in that sequence.
\end{enumerate}
The above results have to be compared with the observation:
The upper limits for black hole masses at
$R = 5\,\kpc$ for the ten model clusters are the strictest
constraints. Since our model clusters all have central densities
around $~5 \times 10^3\:\msun{\rm pc}^{-3}$ and the central density is
directly proportional to the total mass and thus to the depth of the
potential well of the cluster it  influences the binding
energy of the cluster and its stability. One therefore has to compare
the model clusters with Galactic globular clusters in the
 same density range ($\rho_0 = 10^3 $ -- $10^4\:
\msun{\rm pc}^{-3}$), in order to obtain meaningful results.
This is depicted in Fig.~\ref{final-limit}. Analogous to
Fig.\ref{fig-conc-mass}, it shows our ten model
clusters (shaded circles) and the $M_{\rm GC}$ --  $c$ distribution of
the Galactic
globulars in the appropriate density range. The four thick gray lines
confine the region of too low cluster survival rate. Globular
clusters with masses  $M_{\rm GC}$ and concentrations $c$ to the left
of these lines have a probability of less than 1\% ($\eta_1 < 0.01$)
to survive for $10^{10}$ years in a Galactic dark halo consisting of
black holes with masses $M_{\rm BH} = 5 \times 10^4\,\msun$, $10^5\,\msun$,
$2.5 \times 10^5\,\msun$ and $4 \times 10^5\,\msun$,
respectively. These values are obtained from the calculations at
$R = 5\:\rm{kpc}$, since these give the strongest constraints for the
maximum allowed black hole masses. Hence we have to compare them with
the globular clusters at $R \sil 5\:\rm{kpc}$.
 One
sees that all these clusters fall beyond the line
$M_{\rm BH} = 5 \times 10^4 \:\msun$.
Therefore
$5 \times 10^4
\:\msun$ is  an upper limit to the mass of compact dark halo
objects.
This limit is more than
an order of magnitude lower than the value $M_{\rm BH} \simeq
10^6\,\msun$ derived by Lacey~\&~Ostriker~(1985) in order to explain
the disk heating process, that possibly led to the Thick Disk
component in our Galaxy.
Thick Disk formation by heating due to black holes
hence can be ruled out.

In our examination we concentrate on {\em{one}} single effect of
Galactic environment on the evolution of globular clusters, the
influence of massive compact objects building the dark halo of our
Milky Way. It is beyond the scope of this paper to include further
aspects of the dynamical evolution of Galactic globular
clusters, such as disk shocking, influence of the Galactic bulge or
loss of angular momentum of globular cluster orbits due to dynamical
friction, driving them closer to the Galactic center.
Also the role of binaries and mass segregation in the
internal evolution of star clusters are not considered.
Whereas the latter two are negligible for our study, the first two can
be of importance especially in the inner regions of the Milky
Way. Neglecting these corrosive effects when comparing our model with
observed properties of the Galactic population,
leads to a somewhat too large upper mass limit for the black holes.
Thus, a more realistic constraint on compact dark halo objects may
lie below the values derived here.
Concerning globular cluster trajectories,  we assume for
simplicity that they are moving on circular orbits.
To examine the influence of this ansatz, we determine the number
of encounters with compact background objects
 for globular clusters on orbits with different
ellipticities in the Galactic potential,
using the dark matter distribution (equation~\ref{hallali})
derived in Sect.~\ref{discussion-I}. The number $N$ of encounters within a
certain time interval $t$ is $N=\Sigma \cdot n \cdot v t$, with $\Sigma$
being the
collision cross section and $n$ being the number density of background
objects. Calculating $N$ for orbits with the same average
distance $R$ to the Galactic center, i.e. the {\em time} average of $R$ is
kept fixed, but ellipticities ranging from zero to one,
from circular to radial orbits, we
see only a moderate increase of the encounter rate by a factor of
two. This effect gets stronger in a more centrally concentrated dark
matter distribution than adopted in equation~(\ref{hallali})
and vanishes in a homogeneous distribution. We conclude, that using
circular orbits only slightly underestimates the collision rate and
thus only slightly increases the mass limit for black holes in
the Galactic dark halo.

Let us now relax the stringent threshold
of $\eta_1 = 0.01$, adopting for example a value of $\eta_1 = 0.5$,
i.e. if we
require 50\% of the initial clusters to survive. We then get a maximum
permitted  black hole mass which is by a factor of about 5 smaller than
before, thus pushing that value down to $10^4\,\msun$. Since it is
very unlikely, that the halo field stars stem all from disrupted
globular clusters and one can envision many other and more reasonable
formation scenarios, like an intrinsic population of halo stars or
debris from accreted satellite galaxies, the maximum number of disrupted
clusters is more likely in the range of a few tenth of per cent. Then the
 method applied in this paper leads to an upper
limit on $M_{\rm BH}$ of $10^4\,\msun$.

This number has to be compared with limits on $M_{\rm BH}$ derived
by other authors. Moore (1993) obtained values which are
lower by about one magnitude. He quasi-analytically computed the disruption
time
scale as a function of black hole mass for nine halo globular clusters
in our Milky Way. Describing the cluster by a smooth density profile
and thus neglecting the initial velocity dispersion of the cluster
stars, he calculated the energy input per encounter. He then summed up
encounters with randomly distributed impact parameters and velocities
until the total energy input is equal to the binding energy. The
required time interval defines his disruption time $\tau_{\rm D}$. By
adjusting the black hole mass, he fixed $\tau_{\rm D}$ for each
particular cluster to be roughly
half the typical age of a globular cluster  in the Galaxy or about half the
Hubble time: $\tau_{\rm D} = 7 \times 10^9$years.
Thus  he used  only {\em one} encounter history to obtain his upper
limit on $M_{\rm BH}$ for each cluster. Since each encounter history
is dominated by only a few central impacts the scatter between
different encounter histories is huge (see Fig.~\ref{wilder-aff}) and
Moores numbers are plagued by error bars of at least one order of
magnitude. We therefore conclude, that his value of $M_{\rm BH}
\sil 1.1\times
10^3\,\msun$ is still questionable.

A different approach was taken by Rix~\&~Lake~(1993).
They followed the original considerations of Lacey~\&~Ostriker~(1985)
about the effects of massive dark halo objects on galactic  disks,
applying them  to two nearby dwarf galaxies dominated by an extended
dark halo: DDO~154 and GR~8. DDO~154
(similar to DDO~170, Lake~et~al.~1990) results in $M_{\rm BH} \sil 7
\times 10^5\,\msun$ and GR~8 in $M_{\rm BH} \sil 6 \times
10^3\,\msun$.

\section{Conclusion}
\label{conclusions}
The fact that globular cluster with masses in the range $10^4\:\msun
\le M_{\rm GC} \le 10^6\:\msun$ and concentrations as high as $c
\approx 2$ have survived for the past $10^{10}$ years provides strong
constraints on the masses of black holes as possible candidates for
the dark matter in our Galaxy.
 As globular clusters represent only a
small subpopulation of the Galactic halo, a large fraction of them
could indeed have been disrupted by encounters with massive black
holes forming the field star population of the halo,
 with only a few lucky candidates being left behind.
We have calculated  the survival probabilities for different model
clusters in the Milky Way assuming a dark halo consisting entirely of
massive black holes and demonstrated that these probabilities
strongly vary for
 individual encounter histories.
In contrast to former models, we therefore propose that detailed
Monte-Carlo simulations are required in order to determine the maximum
allowed black hole masses. Our calculations lead to an upper mass
limit of $M_{\rm BH} \sil 5 \times 10^4 \:\msun$, which is somewhat
larger than
former estimates but on the other hand small compared to the black
hole masses required to explain the formation of the Thick Disk as the
result of dynamical heating by these massive objects.

The calculations cannot rule out massive black holes
completely. Indeed, black holes with masses of $M_{\rm BH} \sil
5 \times 10^4\: \msun$ could in principle explain the observed upper
concentration limit ($c \approx 2$) of globular clusters as more
concentrated objects would be disrupted. This possibility  however is
still
unlikely given the fact that observations in dwarf galaxies indicate
even smaller upper mass limits.

\newcommand{\AAA}[2]{{\it A\&A\/}, {\bf #1}, #2}
\newcommand{\APJ}[2]{{\it ApJ\/}, {\bf #1}, #2}
\newcommand{\AJ}[2]{{\it AJ\/}, {\bf #1}, #2}
\newcommand{\APJSS}[2]{{\it ApJSS\/}, {\bf #1}, #2}
\newcommand{\Nat}[2]{{\it Nature\/}, {\bf #1}, #2}
\newcommand{\MNRAS}[2]{{\it MNRAS\/}, {\bf #1}, #2}
\newcommand{\CP}[2]{{\it J.\ Comp.\ Phys.\/}, {\bf #1}, #2}
\newcommand{\MAG}[2]{{\it Mitt.~Astron.~Ges.\/}, {\bf #1}, #2}
\newcommand{\APSS}[2]{{\it Astrophys.~Space~Sci.\/}, {\bf #1}, #2}
\newcommand{\et}{{\em et al.}\,\,}

\medskip
\def\rfnce{\par\noindent\hangindent 20pt {}}

\rfnce{%
Allen, C.W.: 1973, {\it Astrophysical Quantities}, University
of London, The Athlone Press, London
}\rfnce{%
Barnes, J.E., Hut, P.: 1986, \Nat {324}{446}
}\rfnce{%
Bronstein, I.N., Semendjajew, K.A.: 1987, {\it Taschenbuch der
Mathematik}, Verlag Harri Deutsch, Thun und Frankfurt~a.~Main
}\rfnce{%
Carney, B.W., Latham, D.W., Laird, J.B.: 1991, \AJ {99}{572}
}\rfnce{%
Carr, B.J.: 1979, \MNRAS {189}{123}
}\rfnce{%
Carr, B.J., Bond, J.R., Arnett, W.D.: 1984, \APJ {277}{445}
}\rfnce{%
Chernoff, D.F., Weinberg, M.D.: 1990, \APJ {351}{121}
}\rfnce{%
Dicke, R., Peebles, P.J.E.: 1968, \APJ {154}{891}
}\rfnce{%
Garret, M.A., Calder, R.J., Porcas, R.W., King, L.J., \hfill Walsh, D.,
Wilkinson, P.N.: 1994, {\em MNRAS} (in prep.)
}\rfnce{%
Hernquist, L.: 1987, \APJSS {64}{715}
}\rfnce{%
Hernquist, L.: 1990, \CP {87}{137}
}\rfnce{%
Hut, P., Rees, M.J.: 1992, \MNRAS {259}{27}
}\rfnce{%
Illingworth, G., Illingworth, W.: 1976, \APJSS {30}{277}
}\rfnce{%
King, I.R.: 1966, \AJ {71}{64}
}\rfnce{%
Lacey, C.G., Ostriker, J.P.: 1985, \APJ {299}{633}
}\rfnce{%
Lake, G., Schommer, R., van~Gorkom, J.: 1990, \AJ {99}{547}
}\rfnce{%
Landau, L.D., Lifschitz, E.M.: 1976, {\it Lehrbuch der
Theoretischen Physik -- Mechanik}, \\ \mbox{Akademie-Verlag}, Berlin
}\rfnce{%
Landau, L.D., Lifschitz, E.M.: 1983, {\it Lehrbuch der
Theoretischen Physik X -- Physikalische Kinetik},Akademie-Verlag,
Berlin
}\rfnce{%
Moore, B.: 1993, \APJ {413}{L93}
}\rfnce{%
Peebles, P.J.E., {\em Principles of Physical Cosmology}, Princeton
University Press, USA
}\rfnce{%
Pfenniger, D., Combes, F., Martinet, L.: 1994a, \AAA {285}{79}
}\rfnce{%
Pfenniger, D., Combes, F.: 1994a, \AAA {285}{94}
}\rfnce{%
Rix, H.-W., Lake, G.: 1993, \APJ {417}{L1}
}\rfnce{%
Rubin, V.C., Burstein, D., Ford, W.K.~Jr., Thonnard, N.:
1985, \APJ {289}{81}
}\rfnce{%
Rubin, V.C.,  Ford, W.K.~Jr., Thonnard, N., Burstein, D.:
1982, \APJ{ 261}{439}
}\rfnce{%
Spitzer, L.: 1958, \APJ {127}{17}
}\rfnce{%
Suntzeff, N.B., Kinman, T.D., Kraft, R.P.: 1990, \APJ {367}{528}
}\rfnce{%
Webbink, R.F.: 1985, in {\it IAU Symposium 113, Dynamics of
Star Clusters}, Ed.~J.~Goodman und P.~Hut, S.541, Reidel
}


\begin{thebibliography}{}
\bibitem[]{} \vspace{-1.0cm}
\end{thebibliography}
\end{document}